# Virtual Reality: Blessings and Risk Assessment

Ayush Sharma[1], Piyush Bajpai[1], Sukhdev Singh[1,2] and Kiran Khatter[1,2]

ayush300897@gmail.com, bajpaipiyush23@gmail.com, sukhdev200@gmail.com, and kirankhatter@gmail.com

[1]Department of Computer Science and Technology
Manav Rachna University, Faridabad-121004, India
[2]Accendere Knowledge Management Services Pvt. Ltd., India

**Abstract**

Objectives: This paper presents an up-to-date overview of research performed in the **V**irtual **R**eality (VR) environment ranging from definitions, its presence in the various fields, and existing market players and their projects in the VR technology. Further an attempt is made to gain an insight on the psychological mechanism underlying experience in using VR device. Methods: Our literature survey is based on the research articles, analysis of the projects of various companies and their findings for different areas of interest. Findings: In our literature survey we observed that the recent advances in virtual reality enabling technologies have led to variety of virtual devices that facilitate people to interact with the digital world. In fact in the past two decades researchers have tried to integrate reality and VR in the form of intuitive computer interface. Improvements: This has led to variety of potential benefits of VR in many applications such as News, Healthcare, Entertainment, Tourism, Military and Defence etc. However despite the extensive research efforts in creating virtual system environments it is yet to become apparent in normal daily life.

## 1. Introduction

The term Virtual in Virtual Reality is defined as "being is essence of effect, but not in fact" and Reality as "something that exists independently in a real state or quality" in the Webster's New Universal Unabridged Dictionary (1989). In general, Virtual Reality means the replication of a real environment and simulation of user's physical presence in this environment using software and hardware [1]. In fact there are different definitions reported by various researchers in the literature. In the book entitled "Understanding Virtual Reality: Interface, Application and design" [2] VR is defined as a medium brought about by technological advances to find practical applications and more effective ways for communication. In book entitled "Virtual Reality Systems" [3] VR is described as the illusion of participation in a synthetic environment rather than the external observation of such environment which has immersive and multi-sensory experience. [4] defined VR in terms of functionality as a real time interactive simulation in which computer graphics is used to create a realistic-looking world and the synthetic world is not static and responds to the user's input. However [5] made an argument that Virtual reality is not related to only technological devices; in fact it is experiencing the physical environment by both automatic and controlled mental processes. Author stated that most of the definitions are device oriented which is not technologically acceptable all the time and addressed this by defining it in a different manner in reference to software developers, media and consumers.

Virtual Reality technology changed the way mathematical simulations and data visualization was perceived. It has also evolved to provide a new level of sophistication in using various sensors such as Visual, Kinesthetic, Auditory etc. It lets the user to sense the presence of the world through computer generated scene by combining various human computer interfaces and sensors. This led the virtual reality as an ideal platform for simulation and decision making that requires frequent and natural interaction with the environment. In fact in the late 1960 and 1970, the panel-mounted display technology was used in the tele-robotic and tele-operations simulations. In his work, Sutherland [6,7] introduced different immersion concepts in simulated environment which are the basis of the current virtual reality technology. Later the research in 1980's by [8], VIVED (Virtual Visual Environment Display) project by NASA in 1984, and then VIEW project (Virtual Interactive Environment Workstation) were the beginning days of the VR technology. The term "Virtual Reality" was actually first introduced by Jaron Lanier, the founder of VPL Research, a company that was involved together with Autodesk Company in NASA's projects [9]. Both the companies first presented the devices with head-mounted displays (HMDs) for interacting with virtual worlds. The HMDs are displays which can be wraparound headset. These displays don't allow light or images from the real world to interfere with the virtual one. With the inception of HMDs, VR started capturing the public imagination. One may further refer to [10] and [11] to gain an insight on virtual environments and the objectives of the NASA's projects, and [12] for VR technological requirements, characteristics and solutions of some more projects using VR in the late 1990.

In literature survey we observed that a number of companies have launched HMDs for VR applications. For an instance the HTC Vive (HTC in partnership with Valve Corporation) has recently gain much popularity through the American Express campaign at the 2015 U.S. Open advertising the VR experience "You vs. Sharapova". The VR experience allows visitors to have virtual experience to play against the famous athlete Maria Sharapova for four minutes by the use of HTC Vive headset with integrated headphones and a special wireless controller used to simulate the tennis swing. Sony's VR headset for the Playstation 4 (Project Morpheus) and Oculus Rift that was acquired by Facebook in 2015 are some other recent examples. Virtual Reality has garnered the attention of researchers and tech enthusiastic, and spanned in almost every field such as Healthcare, Education, Entertainment, Gaming, Tourism, E-commerce etc. In this paper we present the past ground breaking discoveries, present projects and future visions for VR. The second objective of this paper is to give a brief description of all the projects of the leading companies in the VR and mention the risks related to VR technologies. Next section discusses the technology behind VR applications.

## 2. VR Platforms and Toolkits

Looking at the popularity and usage of virtual reality, the need for a user-friendly and cost effective platforms and tools was felt. The aim was to let the users and developers in prototyping, development, testing and debugging of VR applications and immersive systems. This section describes the design of various VR Platforms and Toolkits, and discusses how these help in the creation and execution of immersive applications.

**2.1 VR Platforms**

**Distributed Interactive Virtual Environment (DIVE):** It is a heterogeneous Virtual reality platform works on UNIX and Internet Networking Protocols [13], which provides the VR environment where different users and applications can enter the platform and leave anytime. DIVE hierarchical database consists of entities to describe its data and behaviour. To reduce the latency, some part of database is replicated so that applications can access the DIVE database directly from memory. It uses peer to peer communication and multicast protocol to increase the scalability thus allowing multiple applications in the DIVE world. DIVE system is scalable with respect to number of applications and to the distance between applications on heterogeneous network.

**MASSIVE:** MASSIVE is the "Model, Architecture and System for Spatial Interaction in Virtual Environments", a distributed virtual reality system to support interaction among several concurrent users through graphics, audio and text interfaces. It is based on the spatial model of interaction to facilitate scalability and control the spatial interaction between objects. Each object in a virtual world has an aura to define the extent to which interaction can happen with other objects. When aura of two objects collides, objects become aware of each other and interaction becomes possible by establishing peer to peer connection. Thus scalability depends on the object's aura which limits the number of interactions in the virtual world. However, number of interactions can be increased at the edges of the population. The interaction is controlled by defining the accessibility/awareness level of objects using the concept of nimbus and focus. Focus refers to the extent to which observed object is visible to the observer object, and nimbus describes the observed object's manifestation. Nimbus refers to the space in which an object declares which particular areas of it will be available to others where as object's focus describes its allocation of attention [14].

**SPLINE:** The Scalable Platform for Large Interactive Network Environments (SPLINE) uses peer to peer communication and follows hybrid client-server and multicast approach to support users with low speed link. Spline's world model is an object oriented database to specify the object's behaviour how it will look like and what sound it will make. The lifetime of an object is limited to the application it owns and it exits as long as the application is running. However to make the objects persistent, applications can implement the persistent process to store infrequently visited Spline objects. Spilne breaks the world model into small locales to establish communication among users depending upon their proximity to the locale they are interested in. Thus it is scalable on the basis on number of users in one locale rather than on the number of users in the Spline world [15].

**NPSNET:** It is a 3D visual simulation system developed by researchers of the Graphics and Video Laboratory of the Department of Computer Science at the Naval Postgraduate School (NPS) in Monterey, California. Earlier versions (I and II) of NPSNET implemented Ethernet and ASCII-encoded application level protocol for setting the simulation environment. In these versions, interaction was possible over local network. Then another version "NPSStealth" of NPSNET came into existence which enabled the interaction possible for local and long-haul networks by incorporating SIMNET protocol. With this version, NPSNET touched distributed simulations then later on efforts geared towards large scale simulations. This led to the development of NPSNET-IV which can be used as a simulator for air, ground, nautical (surface or submersible), virtual vehicle, or human [16].

**2.2 VR Toolkits**

**VR Juggler:** It is an open source Virtual Reality toolkit developed by [17] at Iowa State University's Virtual Reality Applications Centre. It was released under GNU LGPL and has produced community-oriented virtual reality application development framework. Since its inception, VR Juggler has been designed to be dynamic system to facilitate run-time re-configurability. The system evolved with numerous changes to facilitate the addition of new functionality and modification in the existing services without affecting the entire system. The subsystem "microkernel" manages all the run-time activities within the system and supports all the services needed for input, configuration and display settings for developing the VR application. Microkernel is a modular architecture that allows loading of those modules which are required by currently running application, and also supports addition, removal and reconfiguration of internal and external managers at run-time. Another subsystem of VRJuggler "Virtual Platform" which supports "write once, run anywhere" slogan. This means VRJuggler application can be created on local VR system and can run on any other VR system. This platform is independent of underlying hardware, architecture and operating system.

**Avocado:** Avocado presents the flexible programming framework for larger-scale collaboration in virtual environments. It is an object oriented programming framework for the development of distributed applications which free the programmers from setting the infrastructure needed to make a VR application distributed. Since the concept of distribution is an integral part of Avocado, the application programmer can focus on the representation of geometry objects and interaction of these objects with the real world by using Nodes and Sensors classes respectively. Avocado uses C++ API to define these objects and Scheme API to create high-level Avocado objects. Use of Scheme scripts allow the testing and debugging during the run time of applications thus provides an environment of rapid prototyping of applications [18].

**GetReal3D:** getReal3D for Unity is a software plug-in launched by Mechdyne Corporation that brings 3D and viewer-based perception to various immersive 3D displays. This plug-in makes the Unity game engine compatible with other immersive 3D environments such as CAVE (Cave Automatic Virtual Environments), Multi-screen and Motion Tracking System. There are four components of getReal3D toolkit: configuration file, daemon, launcher, and plugin. The configuration file configures the settings for display system, hardware and input devices. The getReal3D daemon runs in the background to support the game in running mode. The launcher allows the user to run and deploy Unity games, and plugin component provides the run-time VR behaviors. It allows the users to move through a three-dimensional object through a wide variety of situational simulations. It also helps to control the window creation and corresponding representation from within Unity- 3D. One may further refer to [19] for the detailed study.

**MiddleVR:** MiddleVR, developed in C++, is a graphical configuration tool to setup the virtual reality systems independent of specific software. After compilation of MiddleVR-enabled Unity application, it can run on any VR system supported by MiddleVR. The virtual reality systems it supports are, Head mounted displays, Immersive cubes, Holostages, Holobenches, Powerwalls and 3DTVs. This tool handles the synchronization of computer clusters, S3D stereoscopic displays and menu structure. It also supports the dynamic adaptation of one application from one VR system to another VR system. Various interaction devices such as Kinect, Leap Motion, Oculus Rift, SpaceMouse, Razer Hydra, TrackIR etc. can be managed with the help of Middleware. The motivation behind this tool was to create a platform where user can experience the 3D immersive environment without the extensive VR-specific expertise [20].

**RUIS:** Reality-based User Interface System (RUIS) is an open source toolkit to create an immersive VR application with spatial interaction and stero3D Graphics while interacting with the Unity Scene by incorporating Kinect, PSMove, Razer Hydra and Oculus Rift. It also allows the simultaneous use of these devices in a shared coordinate system. Novice developers can also use this tool to create VR applications with relatively little effort. Once the application is created in RUIS, it can be easily exported on a different computer. This tool comes in two variants: RUIS for processing and RUIS for Unity. Former provides the code templates to help the developers to create their VR applications. But it doesn't provide a thorough 3D programming library to support real-time 3D graphics. Moreover, it was very slow in representing the 3D Graphics. Later, RUIS for Unity came into picture aimed at building more immersive 3D interfaces and virtual systems by porting RUIS to Unity 3D, which is very fast at rendering graphics [21].

**VITAKI:** VITAKI is a vibrotactile prototyping toolkit for virtual reality and video games. It was designed to facilitate the haptic feedback demands of the user to enhance the immersion of the VR environment. Apart from the standard hearing and viewing of a VR environment, VITAKI offers the consumer with features of touch, feel and manipulation of objects present inside it to fulfil the objective of virtual immersion. It has an adaptable hepatic display which can be used in different scenarios [22].

All the VR platforms and toolkits aim to bring a level of realism by providing sensory fidelity which led to the adoption of VR technologies in various fields such as Healthcare, Education, Tourism, Entertainment etc. which is discussed in detailed in the next section.

## 3. Applications of VR

In this section we discuss the applications of the VR technology in various fields. We present some ground breaking discoveries, recent work and future vision in all the fields.

### 3.1 Healthcare

Virtual Reality was introduced in the Healthcare industry soon after its blooming days. In their work [23] discussed the use of VR to treat the phobia of spiders. Authors found that the results were satisfactory with almost 83% patients showing significant improvement with decreased anxiety and fear just after going through an average of four 1-hr sessions only. This technology can also help to train the physicians for emergency medicine as the traditional training involves practicing diagnostic and procedural skills on live patients, see [24]. According to [25], the health sector changes in every 6-8 years with new medical procedures emerging every day. Authors suggested that VR could be a promising area with a high potential of enhancing the training of healthcare professionals citing the fact that the average physician practices 30 years and the average nurse 40 years. Therefore the technology can be used for telesurgical applications to create interactive simulations of the human body. Other interesting applications include the development of immersive 3D environments for the treatment of mental disorders to support psychiatrists and psychologists. [26] also discussed that VR is more likely to be successful if it is systematically integrated into an education and training program which objectively assesses technical skills improvement and increases learning experience. Authors also suggested that the training session should contain of regular intervals rather than a single and short period of extensive practice. In their work [27] used VR to treat patients with anxiety disorders, and the results generated a positive sense of mind among the treated persons. Authors also guided therapists at each step of the process. [28] suggests the use of VR to address the impairments, disabilities, and handicaps associated with brain damage rehabilitation. In the field of laparoscopy, VR simulator training can help to gain the skills to treat the patients. In their work [29] discussed the VR implementation procedure to train novice in short time for the surgical operations. Experiment results suggest that the medical procedure benefitted the scene by increasing the performance level of novice. In fact the simulator-trained group reached results which were equivalent to the experience gained by performing 20-50 actual laparoscopic procedures. [30] suggested that VRWii gaming technology instead of using the traditional methods of stroke rehabilitation by making them play Bingo and Jenga cards. The VRWii is a technology where Virtual Reality was coped up with Nintendo Wii, and it represents a safe, feasible, and potentially effective alternative to facilitate rehabilitation therapy and promote motor recovery after stroke. Authors experiment shows a significant improvement in the mean motor functions of the stroke patients by 7 seconds, and the average treatment time was reduced by 24 minutes. In fact VR is providing physicians with a risk free environment for practicing surgical operations [31]. It has benefited the practitioners with an in-depth description rather than the theoretical studies. A recent report also describes the use of VR in the field of healthcare majorly in four areas such as: Relief of the sensation of pain, exposure therapy, a tool to conquer phobias and Virtual Robotic Surgery [32].

### 3.2 Education

In their work [33] stated that teaching students in a simulated environment style will result in a good academic result, and the use of VR in the field of programming using virtual programming will help students to learn and understand the codes more easily than the traditional learning. In the present days the use of VR in education has been increased due to its rich experience providing and offering a better human-computer interaction for the students. [34] designed a virtual reality platform called the Immersive Touch to stimulate tasks related to neurosurgical operations and reduced the supervised training time of novices for such a tough skill set. [35] also described the use of Second Life which is used for research purposes in the Coventry Universities. One may further refer to [36] for a rigorous review from January 2000 to October 2012. Authors suggested that the success of technology design, instructional approach, and learning experiences would be more when AR (Augmented Reality) along with VR is implemented in the classrooms. The AR-based mobile games allow learners to organize, search and evaluate data and information. Authors have mentioned the negative effects of the technology in the classrooms as teachers would have manipulative powers and hence more control over the content in the system. [37] suggest that VR has a very high potential in education by making learning more motivating and engaging as it is based on three basic concepts: Immersion, Interaction, and User involvement. The reason is that all these concepts allow a physical exploration of objects that are not accessible in reality, for example the planets in the Solar System, helping the learners to understand and memorize them better. The decreases in cost of VR devices are also helping the educational VR to more affordable. In 2015 Google started a project called "Expeditions" for their Google Cardboard [38]. The project's aim was to provide VR headsets to schools on loan so that the student's can experience virtual tours of over more than 100 destinations. It comes with a software the teachers could utilize using a tablet. No student would be charged for these virtual trips and hence compensating for the expensive field trips. Now these days VR technology is being used in engineering to build models in which students get to know about prototyping and iteration in which an object is designed and altered.

### 3.3 Tourism

Virtual Reality would be significantly impacting the tourism marketing sector in the near future. In their work [39] studied the behaviour of persons who were planning an island trip and virtually visiting the different places of their interest. The results proved that the persons not only had a better decision making due to the richer information the VR technology provided but also had more realistic experience of their future journeys. [40] believed in a possibility where one could create new touristic areas (in a virtual environment) incorporated in the online communities and websites to help in the travel planning process, and also in consequence create future tourism destinations. The VR technology has increased the satisfactory rates and consumer trust of the public providing more tourism rate growth in almost every tourist destination; see [41] and [42]. In fact the use of VR technology to have a virtual experience of the place that is still not visited by the tourists has a great potential from the business prospective as well. [43] found that incorporating virtual tours and panoramic photos provided psychological relief to people suffering from travel anxiety or sickness. The Louvre Museum in France showed that when people are made to experience a virtual environment of a museum, that virtual tour also increased the tourist interest in visiting that place physically [44]. In our society we always have people who have their last desires to visit a pilgrimage site but are unable to do so due to disability and other diseases. So the VR technology can come to the rescue by providing an immersive experience of these holy places using a simulation to compensate for their inabilities. [45] also mentioned about the concept of replacing the act of physical travelling with a virtual environment. [46] have the view that the use of VR technology will be more cost effective, and can further help the travel agents and travel groups in better understanding of the requirements of a tourist. The fact is that a tourist service cannot be tested in advance; but the VR technology would help in providing additional sensory and visual information than the use of traditional brochures and pamphlets.

### 3.4 Entertainment

As of now the world has already made a wide acceptance for the 3D viewing of movies and documentaries, Virtual Reality has to be the next step in revolutionizing this field. In their work [47] illustrated how implementing serious games using VR technology could help in developing a new depiction of cultural heritage and improve the visual definition of information, interactivity and entertainment. In the recent talk in a TED conference [48] Author spoke about how the viewers can be more involved in the storyline of a video and sympathize more as well. He collaborated with Arcade Fire to develop a website called "The Wilderness Downtown" where a person could enter his address, and then a boy appears who starts running down the road and stops at a place where your home is/was. People who do not live at that place now or have taken refuge somewhere else could virtually experience about the place where they used to grow up. According to [49], IMAX will open six physical venues (theatres) for VR content worldwide. They are planning to open the first theatre in Los Angeles and the second in China. In the theatres VR headsets created a star breeze to provide 210 degrees view of the film. The CEO of IMAX also told that the VR content would be tied to existing movies. According to a report [50], Anand Gandhi, who is leading filmmaker in Indian film industry and director of Ship of Theseus (National Award Winner), said that virtual reality is profound and virtual reality is reality; indeed it is life itself. In fact VR movie watching is a 360-degree immersive experience where the viewer is at the centre of the action. The first virtual reality movie, a Jesus Christ's biopic, made its worldwide debut at the Venice Film Festival. According to a report "Virtual reality and the silver screen: A match made in heaven" [51], Oculus has built its own video watching program as well named

Oculus Cinema. In which viewer can experience it from a perfect middle seat. The movie is played on a simulated screen with the light bouncing off the seats in the audience, just as it would in reality. According to [52], not everyone play games, but majority goes to the cinema and watch TV. Therefore to reach mass adoption VR can be used in the form of entertainment rather than just gaming. Major studios, have experimented with the form through tie-ins to established franchises, from Interstellar to Game of Thrones. According to another report mentioned in [53], Oculus has decided to get into VR film-making by launching the Oculus Story Studio which was announced at Sundance Film Festival. [54] Published a report in 2017 of the music maestro A.R. Rahman's iconic patriotic track "Vande Matram" which was launched in a VR format offering 360 degree immersive view in NDFC Film Bazaar, Goa. Rahman also said that he used to attend VR workshops in Los Angeles and in India before launching the track for the virtual environment. Rahman also showcased a VR film in UN General Assembly hall on August 2016 showcasing the future project of his team.

### 3.5 News

Virtual Reality has spread its legs in the future of news delivering as well. In their work [55] discussed the idea of immersive journalism which meant to create a first person experience while viewing news. Authors also discussed about the demonstration they had conducted in which they gave participants the experience of being present inside an interrogation room. Recently [56] showed that people can develop a whole new level of empathy when they experience news in an alternate perspective rather than just watching it. Author shared her conclusions in a TED event from an experiment she conducted which considered of a man who suffered from a seizure and collapsed on a footpath. This incident was depicted to the public using the VR technology where people expressed an astonishing set of emotions, and were trying to console and help someone who was not actually present there. In fact VR has that power which the viewers can experience by just learning to be in it and also attracts the young viewers to watch the news [57]. The BBC news described the usage of VR as a tool for their reporters in the near future, and believes that it will be helpful to channels in increasing their viewership and deliver news to a greater audience [58]. BBC along with Aardhman Digital further stated their implementation of VR for depicting a dramatic movie based on migrants from Turkey to Greece. Their R&D department's executive producer, Zillah Watson, also talked about how the storytelling, picture quality and direction are being shaped with new dimensions through the VR technology. BBC has also launched a VR talk show: *No Small Talk* produced by VR City and Lyristic [59]. The show is based on a conversation between two inspiring women Cherry Healey and Emma Gannon. According to the producer, VR can enhance a conversation and invites users more to be a part of the discussion. BBC also highlighted an important issue of Human Trafficking with the help of VR [60]. They have launched a 7 minute long video with 360 video productions as it explores virtual reality as a way of connecting with audiences. [61] also discussed how the VR technique works in the audiovisual pieces and changes the viewer level of implication. The trend in journalism changes with the new technologies in the market. According to authors even Innovation Laboratories want to apply VR technology to their audiovisual pieces to make the information more understandable to the viewer. In a recent study, Media giant CNN is making a great effort by launching CNNVR which will provide the users with 360-degree videos natively viewable in CNN's iOS and Android app, making it the third largest VR compatible app [62]. Despite its availability for Smartphone's, desktop users will also be able to check out the content on their 360-compatible browsers. In an article published by [63], Shubhangi Swarup, Executive Editor of ElseVR, told that VR is one of those technologies that gives 360 degree view of any report and brings a lot of questions inside the viewer's mind which can be answered with the help of text mentioned on their website. Thus VR is the medium in journalism where you cannot hide anything.

### 3.6 Design and Assembly

The application of VR has also been in extensive use in the field of designing. In their work [64] designed a Virtual Design Assembly System (VDAS) which allows engineers to design, modify and assemble mechanical products. As a prototype, the system has a knowledge-based library of standard mechanical fastening parts and hence reducing a great amount of time and work at the stage of modelling itself. The VDAS also has a user-friendly interface feature where the product models can be directly inserted or modified in an interactive VR environment. It tells about the geometric information of the product model, the variation information, the assembly match information and function of assembly planning. VR has also been applied for Virtual prototyping to verify assembly and maintenance processes [65]. Authors discussed the CAD-VR data integration to identify the requirements for designing in Virtual Prototyping and assembling. [66] discussed how Virtual Reality can be used in place of traditional designing as conventional methods lack the depth and sense of realism, and require a lot of imagination. Authors further mentioned that VR technology would help the designer and client together, and also proposed a test arrangement system which can reduce the problem of realism. This directly affected the strategies and business benefits of designing and assembling in the future. The use of VR in the field of architecture and designing with the help of multiple case studies has also been discussed [67]. Those organizations which used virtual reality for architectural design and construction of the physical built environment were at large different from the models created for professional use within the project team and supply chain. [68] stated that an Assembly Process requires various of factors such as optimum assembly time and sequence, tooling and fixture requirements, ergonomics, operator safety, and accessibility, among others. As VR has the potential to integrate human motions into computer aided assembly planning environment, it will result in reduced time and cost effectiveness. According to [69], a construction company, DIRTT has been using Oculus VR to show their clients how their office, home would look after construction. They have designed a wireless set-up that has a backpack and 'Polhemus' technology included. This technology uses an alternating current (AC) magnetic field to track the movement of the user. Recently [70] developed a software called the "Virtual Reality Architecture Walkthrough" where a person can view how the building or the structure would appear in the future from both inside and outside. This not only increased the customer trust but also helped the constructors finding faults and making variations accordingly quite easily. The usage of VR in the field of engineering and construction, with the aim to create a remarkable evolution by giving a whole new experience to a person in visualizing a 3D simulated structure has been discussed by [71]. Earlier it was very expensive to viewing a particular complex than the current economical feature of the VR technology. The technology can be used in showing the different stages of construction which denotes a great potential for the Architecture, Engineering and Construction industry, and hence a person can experience realistic situations without having to care about injuries.

### 3.7 Military and Defence

In their work [72] suggested ways to implement VR technology through remote sensors, intelligent systems and surgical simulations to prepare a combat casualty care for the 21st century. In fact the use of VR technology provides a logical extension of the video camera/television monitor to provide the airplane cockpit's entire field of view. The technology can also be used to train soldiers in a virtual environment by using simulation machines. The major application is in the situations when the area is inaccessible and the operations are required to be conducted. In such cases, the drone or UAV is being viewed with the help of a VR device and perform the required operations. Other uses consist of battlefield simulations and virtual boot camps where the soldiers experienced the consequences and the aftermath of a war. [73] illustrated the use of Virtual Reality Exposure therapy (VRE) in treating active duty soldiers for a variety of anxiety disorders and also posttraumatic stress disorder (PTSD). An immersive simulation of a military convoy in Iraq was conducted for six sessions, and the results had reduced the psychological stress and symptoms of PTSD. [74] motivated their research towards assessing the cognitive functions of the military personnel and train them with stress resilience prior to deployment. [75] evaluated the validity of Virtual Reality Stroop Task (VRST) by making cognitive and neuropsychological assessments to show the impact of environmental stimuli on the performance of military personnel. A recent report by [76] mentioned that the use of VR has been increasing extensively in the form of submarine simulation (simulators mounted onto the hydraulic arms which pitch and roll as a real submarine would, it helps in using the actions of a submarine) and ship simulation. The report also states that Australian Navy is investing multi-million dollars on virtual warships for training purposes. The Naval Trainee has to deal with different situations in that virtual environment for passing the training session. Another blog by [77] reported that VR simulation enables the training of the soldiers without the risk of death or an injury. Other uses include combat visualization to create a depth of illusion and an ideal environment which offers trainee soldiers to experience a particular situation within a controlled area. This heavily utilizes the time and cost spent in military training. An article by [78] mentioned that VR Flight Simulators are taking place of teaching real time flying skills. They have the same structure as that of the real plane which has been mounted on hydraulic lift. The VR simulator tilts, move, twist to change the movements and the system even responds to the user. Thus such procedure also directly affects the cost of the training process and nullifies the damages.

### 3.8 e-Commerce

As we step forward each day in the field of Virtual Reality, another milestone which this technology is about to achieve in the near future is its introduction in the e-Commerce. It would provide the customers with a better visualization and decidability for the products they want to buy rather than the conventional use of just 2D pictures. In their work [79] presented and discussed the ADVIRT which is a first prototype of an adaptive VR store. Introducing VR in e-Commerce websites promises to create a better e-shopping experience which is more natural, attractive, and fun for customers. In another paper [80] discussed the potential advantages of such interfaces and stressed on the need for a better approach to their design and build more usable and effective VR stores. Further [81] attempted to attract the consumer's attention by using VR in Web Stores to enable consumers with a realistic experience of products present over the internet, and thereby tackling the problems associated with consumer's lack of physical contact with products. The study also investigates the circumstances in which VR would enhance a consumer's learning about a product. [82] had shown the use of virtual reality for online shopping environments to offer an advanced customer experience compared to the conventional web stores and further improve the customer trust. The paper presents a prototype of a virtual shopping mall environment, with the proof of an empirical study explaining that a virtual reality shopping environment would be preferred by customers over a conventional web store. Such application would also facilitate the assessment of the e-vendor's trustworthiness. [83] disclosed an online merchandising and promotion system that closely simulated an interior space of an actual place of business. The consumers used avatars to explore the simulated space; they become simultaneously familiar with both the simulated and actual places of business, and thereafter can shop at both locations with equal ease. According to [84] customers still prefer to buy from retail stores because they can feel from all of the five senses for what they are buying instead of using e-Commerce sites because they cannot feel the commodity they are buying. Thus the e-Commerce giants are taking all the steps to increase their customer's trust and satisfaction by integrating virtual reality in a better way. Shopify, a market leader in this field, showcased some of its prototype virtual reality technologies in the yearly Unite Conference. They launched a VR application compatible with most devices and browsers to set a new benchmark for product visualization [85]. A blog published by [86], shares that different companies are continuously taking risks to make a progress in their customer experience development. Recently an Australian retailer, Myer, has partnered with e-Commerce giant eBay in launching a VR department store [87]. This gives the customer an opportunity to feel the product before buying or adding to cart and improves retailing and browsing experiences.

### 3.9 Sports

Introducing Virtual Reality in sports is also one of the main focuses of VR technology, and is already being successfully done by companies like Virtually Live. VR technology can be used to influence the exercise regimes of people, see [88]. Authors used a HMD and showed the subject computer-generated images while the person exerted his force on the machine. As the user exerts forces against the machine, their perception is that as if they are exerting forces against the objects represented in the images. [89] developed a sports training system with the help of VR to analyze and assess the quality, stress levels and the reactions by the athletes in different situations. [90] applied the VR science in developing the skill set of a handball goalkeeper. The goalkeeper was presented a virtual scenery like of a real game to study his motor behaviour and natural gestures when facing a thrower. [91] provided a better view model for the in-depth analysis of the sports players. Video playback doesn't permit such analytical perspective and therefore introducing VR technology for such would provide a better understanding of an athlete's performance. They conducted case studies to showcase how to use information from visual displays to analyze a player's future course of action. [92] in a TED conference discussed how AR and VR can be linked with National Football League (NFL). He explained that if some smart-glasses, just like the Google glass, can be used for their ubiquitous computing feature by installing it on the helmets of the players in the field. Further the viewers can use their VR devices to connect with these glasses, and what they view from their glasses can be streamed for the audience who can use their VR devices and connect to this live stream and watch the game according to their desired player's perspective. He believes that this concept would not only create a new viewer experience in the field of sports but also develop empathy among people as they would be able to realize how difficult it was or not for the player on the field in making certain moves. It will also reduce the risk of match-fixing as the coach and people present in the stadium can watch what the player is actually watching. Recently [93] provided a complete 360 live view of the Rio Olympics held in August 2016 which the users can watch using their VR devices or cardboards. This was a step forward in the usage of virtual reality as a medium of sharing sports events and journalism.

### 3.10    Gaming

During the dawn of virtual reality, the technology was already being remarked as something that would bring a significant change in the field of gaming and set up a higher level to 3D which people used to experience before. [94] discussed how the visual simulation and virtual reality communities provide not only a delivery system for organizational video game instruction and training but also has the potential to affect a greater audience. The feeling of presence in an alternate environment provides better simulation and an interactive perspective of the storyline of the game. NVIDIA technologies such as VR SLI enables us to assign the rendering for each eye to different GPUs and Multi-Res Shading feature renders every part of an image at a resolution that better matches the pixel density of the warped image needed for VR [95]. Traditional games were mostly running at 60 frames per second but since VR has made its benchmark in this field, companies are now developing games which could run over 90 frames per second to meet the immense graphical demands of public. Niantic in July 2016 released a world record breaking game called the Pokémon GO for the Android and iOS devices grossing more than $10 billion. The game was based on the simple basics of AR (Augmented Reality) technology where the user can travel in the physical environment and using their phone's camera and GPS can catch creatures of the virtual environment [96]. The game became very popular around the globe as it was the first multiplayer game which is played outdoors. Now Niantic's next target is to introduce virtual reality in their massive success to engulf the players in an interactive virtual environment of the Pokémon world. eSports is a field which has grown faster than any other sport in the current generations and to meet the audience's interest, Sliver.tv (2016) founded by Mitch Liu in 2015 has developed a platform to record, view, and stream top eSports games in fully immersive, 360° cinematic VR videos which are compatible with every headset [97]. The new cutting edge technology has opened a large amount of viewing of player perspectives for the user. More projects and Players in the gaming industry are discussed in the next section. PlayStation VR and AMD Sulon are some other recent examples which define the advancement in the field of gaming [98]. VR has been adopted fastest and most comfortably in this application than any other field till now.

Success of all these applications depends on how well they meet the requirements in the respective field, simulate the real-world experience and provide the immersive virtual environment to the user. Attention around VR seems to be reaching very high which is evident by the involvement of major players in the virtual reality market, discussed in the next section.

### 4. Major companies involved in VR

The objective of this section is to provide the insight on the projects of various companies working on VR. Table 1 list all the key players, their products, investors and the fields in which they are involved. From the Table it can be seen that most of VR player belong to US. In 2015 a survey was conducted by MarketsandMarkets.com [99] for the purpose of defining, describing, and forecasting the global virtual reality market between 2016 and 2022. The findings of the survey suggest that the virtual reality market is estimated to grow from USD 1.37 Billion to USD 33.90 Billion by 2022, at a CAGR of 57.8% between 2016 and 2022. The VR market for software components is also expected to grow at the highest rate due to the increasing adoption of VR software platforms and applications across the globe. However the VR market is growing rapidly but still there are many risks involved in this technology which are discussed in next section.

### 5. Risks involved with VR

### 5.1 Physical risks

Virtual Reality, since its introduction in the industry, has been questioned for its physical impact on a person's health and mental state of mind. Many devices have not perfected their head-tracking systems yet and such discrepancies could lead to motion sickness or nausea [100]. The U.S. Army has categorized the usage of the Oculus Rift risky as they do not want to introduce a thing which is not perfect yet for the soldiers. Oculus has also warned its users in its terms and policies quoting "Some individuals may also experience severe dizziness, epileptic seizures or blackouts when exposed to certain flashing lights or patterns" [101]. Although VR exposure therapy is in wide use by the society but the sudden confrontation of fear by the patient who is not ready yet may also lead to a severe shock. Samsung has also released an official statement that such devices must not be used more than 5-6 hours a day and there must be a break of at least 10 minutes after every 30 minutes of VR exposure [102]. Such measures are required to reduce the cases of reduced coordination and headaches. [103] in March 2016 reported the dangers of virtual reality through the malfunction of camera sensors in HTC Vive devices which almost lead to people colliding with an obstacle or wall.

### 5.2. Privacy risks

Information and data security is yet another hurdle the VR technology faces in its path to success. Most of the platforms are Open Source and this allows VR developers from around the world to publish and share their content. ProtoSphere (developed in November 2012) is one of these open source platforms which has been highly integrated with Microsoft's communication system [104]. All types of communications including voice, speech/gesture, presence and instant messages pass through the Microsoft's Lync server and hence giving rise to the risk of information leakage. Julie LeMoine, the CEO of HorizonIR also stated risks with less level of occurrence where a user can make modifications in their avatar or the virtual environment present around him leading to people seeing what the product wasn't meant to show [105]. Therefore, according to her, the companies must set limitations to the customization of an avatar and what all changes a person can or cannot make when present inside a virtual environment. According to [106], Facebook terms and policies allow them to keep and share a lot of information about a person [107]. When Facebook acquired Oculus VR, a gaming startup, it highlighted yet another risk of privacy as Facebook could manipulate the conditions of the policy. It could track almost every behavior of the person and change the elements of virtual environment as well. Therefore, a company must not alter a customer's privacy guidelines.

### 5.3. Behavioural risks

Studies have shown that Virtual Reality may also affect the behaviour of a person to some extent. Telis, a Canadian telecommunications company, uses the AvayaLive Engage for recruiting employees and board meetings, see [108]. According to Paul McDonagh-Smith, who is a learning practice leader of Avaya, "It is almost like a mask they can sit behind". [109] contradicted the above statement in their article "Real Risks of Virtual Reality" by mentioning that people could also have a dark side which their avatars would not reveal. The avatars in the virtual environment do not express facial expressions, body language, eye contact, etc. in detail. Therefore, a virtual environment could abstract a user's behaviour as the interaction may seem real and personal but it also offers the same potential for rudeness, dominance and harassment just like in any other means of communication.

### 5.4. Investment risks

As the Virtual Reality technology is still in its early stages, gaining funds from the investors is one of the most major problems faced. Even if the product or project is capable of giving a good outcome, one can never predict the level of public acceptance it will get and therefore unlike Facebook, who invested $2 Billion in this industry [110], other companies such as IMAX, Audi and Pepsi seem to make cautious steps since Virtual Reality's downfall from 2014. Anders Gronstedt, President of the Gronstedt Group, Inc. quoted that "The VR accessories are more suited for gaming than of business. There are a lot of issues that need to be worked out", see [111]. Reports by SuperData Research show that Facebook sold nearly 355,000 VR headsets while the threshold expected by their CEO Mark Zuckerberg was between 50 million to 100 million VR headsets. Facebook has also slashed the costs of their VR headsets due to Sony PlayStation VR and HTC Vive taking the lead in the market [112]. According to the article published by The New York Times, Facebook has decided to invest $3 billion more over the next decade on virtual reality projects [113]. Although Steam has also made large investments, Valve's founder Gabe Newell has stated that he is "pretty comfortable with the idea that it will turn out to be a complete failure", see [114]. Nate Mitchell, the founder of Oculus, has been advocating Facebook and attracting investors by saying "We have always tried to set the expectation that VR is a very, very long-term play" in an interview with [115]. [116] shared the news that the AvayaLive Engage which was one the three major enterprise-based Virtual Environment platforms has been shut down because the required plugins were no more supported by the browser and it was cost prohibitive as well. Microsoft's HoloLens is also expected to meet the same fate and result in a failure just like Google Glass, in reference to a report published by [117]. Such cases make an investor question the future scope of investment in the virtual reality industry.

## 5. Conclusion

Virtual reality is a growing industry that not only serves as a profit but also as a future technology which would be helping the human kind. The technology has led to VR devices being improved to a large extent in consideration of the user's health and has set guidelines for its optimal use. The main risk now being that humanity might lose the human connection we have in between because of such virtual means of communication. With VR we are entering a new era of a nascent stage technology, and it is hard to predict rather we would make the jump or fall into an abyss but as we all say, "Without risk, there is no reward". VR is like a child with a little knowledge who loves taking risks, has downfalls but as a parent (users and developers of VR) we are the ones who have to encourage it to get back up again and therefore make it successful when it turns an adult.

**Acknowledgement**
Authors express their sincere gratitude to the Research Mentors of Accendere Knowledge Management Services Pvt. Ltd. Although any errors are our own, and it should not tarnish the reputations of these esteemed persons.

**Table 1:** Details of the companies in VR

| Name of the Company and Website | Founded By | Headquarters | Projects/Products | Leading Investors | Fields involved |
|---|---|---|---|---|---|
| Google (vr.google.com) | Larry Page, Sergey Brin | Googleplex, California, U.S. | Daydream, cardboard and Google Glass | | Gaming, Entertainment, Healthcare, Sports, Videography, Tourism, Google Earth |
| Facebook/Oculus (oculus.com/rift) (facebook360.fb.com) | Palmer Luckey, Nate Mitchell (Oculus) Mark Zuckerberg | Menlo Park, California, U.S. | Oculus Rift, Facebook 360 | Samsung, Spark Capital, Andreessen Horowitz | Gaming, Tourism, News, Entertainment, Military, Architecture, Art |
| WorldViz (worldviz.com) | Andrew Beall, Peter Schlueer | Santa Barbara, California, U.S. | Vizard VR Toolkit, PPT | Intel Capital | Building Architecure, Design and Assembly |
| Bricks and Goggles (bricksandgoggles.com) | Ingmar Vroege, Gertjan Leemans | Nieuwendijk, The Netherlands | Luxury Apartment by Bricks&Goggles | | Home decoration, Building Architecture, Design and Assembly |
| Marxent Labs (marxentlabs.com) | Barry Besecker, Beck Besecker | Dayton, Ohio, U.S. | VisualCommerce | Techstars, Detroit Venture Partners | Online shopping, eCommerce |
| Unity Technologies (unity3d.com) | Nicholas Francis, Joachim Ante | San Francisco, California, U.S. | UnityVR | China Investment Corporation, iGlobe Partners, Thrive Capital | Gaming, Web Engines, Entertainment |
| Microsoft (microsoft.com/microsoft-hololens) | Bill Gates, Paul Allen | Redmond, Washington, U.S. | HoloLens | | Entertainment, Education, News |
| Magic Leap (magicleap.com) | Rony Abovitz, Sam Miller | Plantation, Florida, U.S. | Magic Leap One | Google, Qualcomm, Alibaba Group | Almost in all the fields discussed in the previous section |
| Vuzix (vuzix.com) | Paul Travers | Rochester, New York, U.S. | M100 and M300 smart-glasses | Intel Capital, Lenovo | Entertainment, Communication, Gaming |
| castAR (castar.com) | Jeri Ellsworth, Rick Johnson | Mountain View, California, U.S. | CastAR | Playground Capital | Gaming, Entertainment |
| Sixense (sixense.com) | Amir Rubin | Los Gatos, California, U.S. | STEM System | Kickstarter | Gaming, Entertainment |
| Retinad Analytics (retinad.io) | Kevin Ouellet, Samuel | Ontario, Canada | Analytics Platform for 360VR media | PasswordBox, Zynga, Colopl VRFund | e-Commerce, Business Analytics |

| | Poirier | | | | |
|---|---|---|---|---|---|
| Virtually Live (virtuallylive.com) | Jamid El-Imad, Jesus Hormigo | San Francisco, California, U.S. | VirtuallyLive App | Tectus Group, W Investment | Sports, Entertainment |
| Vega (vega25.com) | Tejhaswi Bharadwaj | San Jose, California, U.S. | Zenie App | | Home Decoration, Design and Assembly |
| byondVR (byondvr.com) | Noam Levavi, Eran Galil | New York, U.S. | VR Publisher Toolkit | | Sports, Entertainment, Tourism, eCommerce |
| Sulon Technologies (sulon.com) | Dhan Balachand | Markham, Ontario, Canada | Cortex, Sulon Q | AMD | Gaming, Entertainment, Tourism, |
| HTC and Valve (htcvr.com) | Cher Wang, Peter Chou (HTC) Gabe Newell, Mike Harrington (Valve) | Taoyuan, Taiwan (HTC) Bellevue, Washington, U.S. (Valve) | HTC Vive, SteamVR | Aurora AR, VRChat, WEVR, Beats Electronics | Gaming, Entertainment |
| Nvidia (nvidia.com) | Jen-Hsun Huang | Santa Clara, California, U.S. | VRWorks | DARPA, TriplePoint Capital | Gaming, Entertainment |
| Samsung Electronics (samsung.com) | Lee-Byung Chul | Suwon, Ch'ungch'ong-namdo, South Korea | Samsung Gear VR | Facebook, HTC, Oculus | Gaming, Entertainment, Tourism |
| Sony (sony.com) | Masaru Ibaku, Akio Morita | Sony City, Minato, Tokyo | Project Morpheus (PlayStation VR) | | Gaming, Entertainment |